\documentclass[aps,letterpaper,prb,twocolumn,notitlepage,showpacs,amsmath]{revtex4-1}
\usepackage{bbm,color,amsmath,amssymb,graphicx}
\usepackage[pdftex]{hyperref}
\usepackage{braket}
\usepackage{stmaryrd}

\providecommand{\average}[1]{\left\llbracket{#1}\right\rrbracket}

\begin{document}

\title{Decoherence of two coupled singlet-triplet spin qubits}

\author{Yang-Le Wu}
\author{S. Das Sarma}
\affiliation{Condensed Matter Theory Center and Joint Quantum Institute, Department
of Physics, University of Maryland, College Park, Maryland 20742-4111, USA}

\begin{abstract}
We study a pair of capacitively coupled singlet-triplet spin qubits.
We characterize the two-qubit decoherence through two complementary measures, 
the decay time of coupled-qubit oscillations and the fidelity of entangled 
state preparation.
We provide a quantitative map of their dependence on charge noise and field 
noise, and we highlight the magnetic field gradient across each 
singlet-triplet qubit as an effective tool to suppress 
decoherence due to charge noise.
\end{abstract}

\date{\today}
\maketitle

\section{Introduction}

Localized electron spins in semiconductor quantum dots provide a promising 
architecture for quantum computation~\cite{Loss98,DiVincenzo00,Levy02,Koppens06}.
Compared with alternative platforms (e.g., trapped ions, trapped atoms, or 
superconducting circuits),
semiconductor spin qubits have the threefold advantage of (1) fast single-qubit gate 
operations, (2) long coherence time~\cite{Veldhorst14},
and (3) potential for scalability~\cite{Friesen03,Taylor05} because of advanced 
fabrication technology developed for semiconductor integrated circuits~\cite{Zwanenburg13:QE}.
There has been significant progress in the experimental development of 
high-fidelity single spin qubit gate operations in different platforms using 
different materials.
Unfortunately, despite considerable experimental 
efforts~\cite{vanWeperen11,Nowack11,Shulman12,Veldhorst15,Nichol17}, it remains 
challenging to engineer high-fidelity two-qubit entangling gates for semiconductor spin qubits, 
as decoherence due to environmental noises is exacerbated by 
comparatively weak coupling between localized electron spins.
In fact, the experimental progress in developing two-qubit gates in 
semiconductor spin quantum computing platforms has been disappointing so far 
when compared with the corresponding situation in superconducting and ion trap 
platforms.
The relatively low entangling-gate fidelity of spin qubits
($0.9$ in GaAs as reported in Ref.~\onlinecite{Nichol17})
is currently the main obstacle to unlocking their aforementioned advantage as 
a platform for large-scale quantum computing,
and it calls for a better understanding of the effect of environmental noises on 
the dynamics of two coupled semiconductor spin qubits in order to enhance the 
coupled-qubit fidelity.

In this paper, we undertake this challenge and study the decoherence of two coupled 
singlet-triplet spin qubits~\cite{Levy02,Petta05},
which are among the actively studied spin qubits with the added advantage that 
two-qubit gate operations have been demonstrated in GaAs-based singlet-triplet 
qubits~\cite{Shulman12,Nichol17}.
Each singlet-triplet qubit consists of a pair of exchange-coupled electron 
spins localized in a double quantum dot,
and the two qubits interact via an Ising-type capacitive coupling~\cite{Shulman12}.
Compared with the exchange-coupled spin qubits~\cite{Loss98} studied in a 
previous paper~\cite{Throckmorton17},
where two localized electron spins are coupled through the Heisenberg 
coupling~\cite{Scarola05,Hu00},
the singlet-triplet system we consider here
enjoys full two-axis control through purely electrical gating~\cite{Wu14}
and is protected from homogeneous magnetic field fluctuations in 
each double quantum dot~\cite{Levy02}.
It also operates in a larger active Hilbert space due to the lack of spin 
conservation, and has more complicated dynamics and richer physics.
This makes it harder to extract useful insights from analytical solutions~\cite{Wang15}.
Instead, we study the coupled singlet-triplet qubits through numerical calculations
in order to provide quantitative insight into the detrimental role of 
(electric) charge and (magnetic) field noise on the two-qubit Ising gate 
operations.
We mention that sophisticated dynamical decoupling schemes have already been 
developed for semiconductor singlet-triplet qubits enabling efficient and 
fault-tolerant gate operations~\cite{Wang12,Wang14,Zhang17,Throckmorton1709},
and substantial progress is likely in the near future once two-qubit 
entangling gates achieve higher fidelity, making our current theoretical 
analysis timely.

We consider coupled-qubit decoherence from two different types of 
environmental noises,
charge noise from charge fluctuations in each qubit device~\cite{Hu06},
and field (Overhauser) noise due to nuclear spin dynamics in the 
semiconductor background~\cite{deSousa03}.
We assume that the noises are slow relative to the qubit dynamics, and 
we model them in the quasistatic bath approximation by averaging observables over 
time-independent but randomly distributed disorder configurations.
The quasistatic bath approximation has been used extensively in the 
semiconductor spin qubit studies and is generally considered to be quite valid 
in most situations~\cite{Throckmorton17}.

The decoherence of the coupled qubits is examined quantitatively through a 
pair of complementary probes.
First, we extract a characteristic time scale, the two-qubit coherence time, 
from the envelope decay of the coupled-qubit oscillations.
This measures the persistence of the initial state information in the presence 
of environmental noises.
It also provides a direct physical measure of the time duration of coherent 
gate oscillations.
Second, we compute the fidelity of preparing an entangled state through time 
evolution from an unentangled product state.
This quantifies the ability of the coupled qubits to carry out a precise unitary 
transformation despite the fluctuations in coupling parameters, and serves as 
a simple proxy for gate fidelity.
We also note in this context the interesting possibility of singlet-triplet 
semiconductor qubits being effective quantum sensors because of their delicate 
dependence on charge and field noises.  In particular, the type of theoretical 
analysis presented in the current work can be inverted and used for a 
quantitative determination of background charge and/or field fluctuations from 
the singlet-triplet entanglement information as described in our study.

We study the dependence of these two fidelity measures on charge noise, 
field noise, as well as the magnetic field gradient across each 
singlet-triplet qubit.
When the average magnetic field gradient is zero, we find that the coupled 
singlet-triplet qubits are significantly more susceptible to charge noise than 
field noise.
This difference is manifested in both the two-qubit coherence time and 
the entanglement fidelity, although less pronounced in the 
latter.
The situation changes dramatically when a strong magnetic field gradient is 
applied in each qubit.
As the magnetic field gradient increases, the coupled-qubit system becomes more 
sensitive to field noise and less to charge noise.
In the regime dominated by the magnetic field gradient, charge noise becomes 
relatively inconsequential and the decoherence of the coupled qubits is mainly 
driven by field noise, in sharp contrast to the situation without a strong 
magnetic field gradient.
The change of noise sensitivity driven by the magnetic field gradient is a unique 
feature of the singlet-triplet system and has no direct counterpart in a 
system of two exchange-only qubits.
We mention that the magnetic field gradient induced strong suppression of the 
charge noise effect on the two-qubit Ising gate operations for singlet-triplet 
qubits provides encouraging prospects for Si-based quantum computing platforms 
since isotopic purification enables the elimination of the nuclear field noise 
in Si systems~\cite{Witzel10}.

The paper is organized as follows.
In Sec.~\ref{sec:t2}, we describe the Ising model of two single-triplet spin 
qubits and discuss the two-qubit coherence time $T_2^*$.
We provide an operational definition of $T_2^*$ and examine its dependence on 
both charge noise and field noise, as well as the magnetic field gradient.
In Sec.~\ref{sec:ent}, we introduce the fidelity of entangled state 
preparation and examine its parametric dependence.
In Sec.~\ref{sec:conclusion}, we summarize our numerical results and highlight 
the implications for future experiments.

\section{Two-qubit coherence time}
\label{sec:t2}

The system of two capacitively coupled singlet-triplet spin qubits
is described by the following Ising-type Hamiltonian~\cite{Shulman12,Nichol17}
\begin{equation}\label{eq:h}
H=\varepsilon J_1J_2\,\sigma_1^z\sigma^z_2
+J_1\sigma_1^z+J_2\sigma_2^z
+h_1\sigma^x_1+h_2\sigma^x_2.
\end{equation}
Here we work in the singlet-triplet basis, with the $\sigma^z_i=+1$ ($-1$)
eigenstate denoting the singlet (triplet) state of the two electrons in the 
double quantum dot constituting the $i$th spin qubit ($i=1,2$), respectively.
For each spin qubit, the Zeeman $h_i\sigma_i^x$ term is controlled by the 
magnetic field gradient $h_i$ across the corresponding double quantum dot,
and the $J_i\sigma_i^z$ term is controlled by the intraqubit exchange 
coupling $J_i$ between the two electrons in the qubit.
The coupling $\varepsilon J_1J_2 \sigma_1^z\sigma_2^z$ between the two qubits
comes from the capacitive dipole-dipole interaction, with a strength approximately
proportional to the product of intraqubit exchange couplings $J_1J_2$ as 
argued empirically in Ref.~\onlinecite{Shulman12}.

We employ the quasistatic bath approximation and model the environmental 
noises by averaging observables over time-independent but randomly 
distributed model parameters.
Specifically, the couplings $J_1$ and $J_2$ are drawn from a Gaussian 
distribution with mean $J_0$ and variance $\sigma_J^2$ but restricted to 
non-negative values,
and the transverse fields $h_1$ and $h_2$ are drawn from a Gaussian 
distribution with mean $h_0$ and variance $\sigma_h^2$.
Physically, the parameter $J_0$ is the (average) exchange coupling between the 
two electrons in each double quantum dot, 
and the parameter $h_0$ is the (average) magnetic field gradient applied 
between the two electrons.
In a typical experiment~\cite{Martins16,Nichol17}, the intraqubit exchange 
$J_0$ is on the order of $10^2$MHz, and the (quasistatic equivalent of) charge 
noise $\sigma_J$ is on the order of $10^{-2}\sim 10^{-1}J_0$.
The field noise $\sigma_h$ may range from up to $J_0$ in GaAs
to essentially negligible in isotopically purified $^{28}$Si.

For our numerical calculations,
we fix the interqubit coupling parameter $\varepsilon$ to $0.1J_0^{-1}$, and 
focus on the effect of the remaining dimensionless parameters $\sigma_J/J_0$, 
$\sigma_h/J_0$, and $h_0/J_0$.
For the disorder average, we typically use a sample size between $10^4$ and 
$10^5$ for each parameter set.
We note that the dependence on $h_0$ is an important new 
element in the physics of Ising-coupled singlet-triplet qubits with no analog 
in the corresponding exchange coupled spin qubits studied in 
Ref.~\onlinecite{Throckmorton17}.

\subsection{Decay of coupled-qubit oscillations}

\begin{figure}[]
\centering
\includegraphics{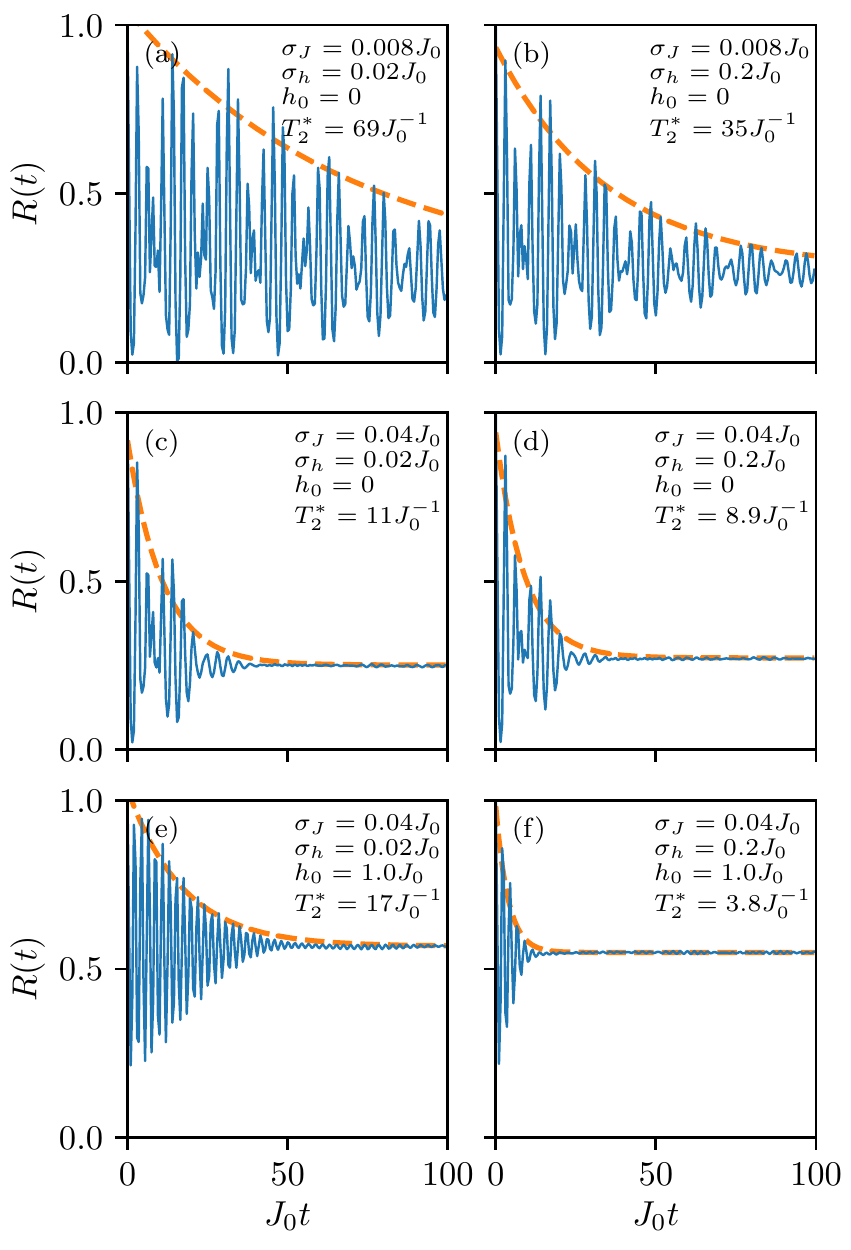}%
\caption{\label{fig:typical}
Representative examples for the disorder-averaged return probability $R(t)$ 
(blue) for various system parameters.
The dashed orange lines show the least-squares fit of the upper envelope
of $R(t)$ to the exponential decay form in Eq.~\eqref{eq:decay-ansatz}.
The system parameters $\sigma_J,\sigma_h,h_0$ and the extracted coherence time 
$T_2^*$ are listed in each panel.
}
\end{figure}

In the following we introduce an operational definition for the 
two-qubit coherence time $T_2^*$ from the envelope decay of the coupled-qubit 
oscillations.
Without loss of generality, we consider the product initial state 
$\ket{\psi(0)}=\ket{\uparrow}_x\otimes\ket{\uparrow}_x$ of the coupled qubits,
and we compute the dynamics of the disorder-averaged return probability to the 
initial state
\begin{equation}
R(t)=\average{\big|\braket{\psi(0)|\psi(t)}\big|^2\,}.
\end{equation}
Here, the double bracket denotes the average over disorder realizations.
As the initial state $\ket{\psi(0)}$ is not an eigenstate of the coupled 
qubits, the return probability is oscillatory in time.
The oscillations have a typical frequency on the order of $J_0$,
driven by the intraqubit exchange coupling $J_i\sigma_i^z$ terms in the 
Hamiltonian [Eq.~\eqref{eq:h}],
and they are further modulated by beats with frequency around $\varepsilon J_0^2$ 
due to the interqubit coupling $\varepsilon J_1J_2\sigma_1^z\sigma_2^z$ term.
The oscillations in $R(t)$ are damped by environmental noises through disorder 
averaging,
in a fashion mathematically similar to (although physically distinct from) the 
decaying Rabi oscillations of a single qubit in the presence of environmental noises.
Very loosely one can think of these oscillations as the two-qubit Rabi 
oscillations decaying due to charge and field noises.
A few representative examples of the decaying $R(t)$ curves are shown in 
Fig.~\ref{fig:typical}, using noise parameters approximately 
consistent with experimental situations.

We extract from $R(t)$ a characteristic time scale $T_2^*$ associated with the 
decay of the oscillation envelope of the coupled qubits, and use it as a 
quantified measure of the coupled-qubit decoherence.
Compared with the exchange-only case~\cite{Throckmorton17}, the $R(t)$ 
oscillations here have more 
complicated wave forms, with extra beats in the decay envelope.
Since these additional features are irrelevant to our main 
goal of characterizing the damping effect of environmental noises,
we disregard them and adopt a simple fitting procedure that focuses only on the 
decay envelope.
Operationally, we take the upper envelope of the $R(t)$ oscillations and 
perform on it a least-squares fit to an exponential decay of the form
\begin{equation}\label{eq:decay-ansatz}
R(\infty)+A\,e^{-t/T_2^*},
\end{equation}
where $R(\infty)$ is estimated from the asymptotic value of $R(t)$ and $A$ is 
a nuisance parameter of no interest in the current work.
The fitted upper envelope of $R(t)$ and the coherence time $T_2^*$ are shown 
in Fig.~\ref{fig:typical} for a few representative examples.

Compared with the exchange-only case studied in 
Ref.~\onlinecite{Throckmorton17}, here we are using a slightly different 
operational definition for the two-qubit coherence time.
This is necessitated by the irregular wave forms of the coupled-qubit 
oscillations allowed by a larger Hilbert space.
We emphasize that this alternative choice only introduces moderate variations 
in the numerical value of $T_2^*$ and does not affect our conclusions 
qualitatively.
In addition, it is worth emphasizing that the coherence time in this paper 
measures the decay rate of the \emph{oscillation envelope} of the 
disorder-averaged return probability $R(t)$, rather than the decay rate of 
$R(t)$ itself. As we noted in a previous paper~\cite{Wu17}, the latter 
definition is more appropriate for a large number of coupled qubits, whereas 
the definition adopted here provides a more precise measure of the decoherence 
process within a low-dimensional Hilbert space appropriate for just two coupled qubits.

\subsection{Quality factor}

The two-qubit coherence time $T_2^*$ as defined above measures the time it 
takes for the envelope of the damped oscillations in $R(t)$
to decay to $1/e$ of its initial value.
Hence, the dimensionless combination $J_0T_2^*$ can be thought of as the number of 
appreciable oscillations in $R(t)$ before it saturates to the asymptotic value $R(\infty)$.
In the results presented in Fig.~\ref{fig:typical}, the dimensionless
parameter $J_0T_2^*$ varies from $69$ [Fig.~\ref{fig:typical}(a)] to
$4$ [Fig.~\ref{fig:typical}(f)], with the results of Fig.~\ref{fig:typical}(f) 
being the most representative of the current experimental state of 
the arts in GaAs singlet-triplet qubits~\cite{Shulman12,Nichol17}
where only a few ($\lesssim 5$) two-qubit gate oscillations have so far been achieved 
experimentally.
(We mention, however, that the experiments~\cite{Shulman12,Nichol17} are 
mostly in the $h_0>0$ regime more appropriate for the discussion in the next 
subsection of this paper.)
To convert this into a number with a normalization comparable with other 
fidelity measures, we further define the quality factor~\cite{Throckmorton17}
\begin{equation}
Q=\exp\left(-\frac{1}{J_0T_2^*}\right).
\end{equation}
This quantity is essentially the exponential decay factor for the 
return probability oscillation envelope over $\Delta t=1/J_0$,
the intraqubit exchange coupling time scale.

In the rest of this section we present numerical results on the decoherence of two 
singlet-triplet qubits using the quality factor $Q$ as a quantitative measure 
of coherence.
We will make comparisons with the exchange-only case studied in 
Ref.~\onlinecite{Throckmorton17} where appropriate, and explain how the 
additional tunability of the singlet-triplet system through the magnetic field 
gradient may be exploited to suppress decoherence due to charge noise.

\begin{figure}[]
\centering
\includegraphics{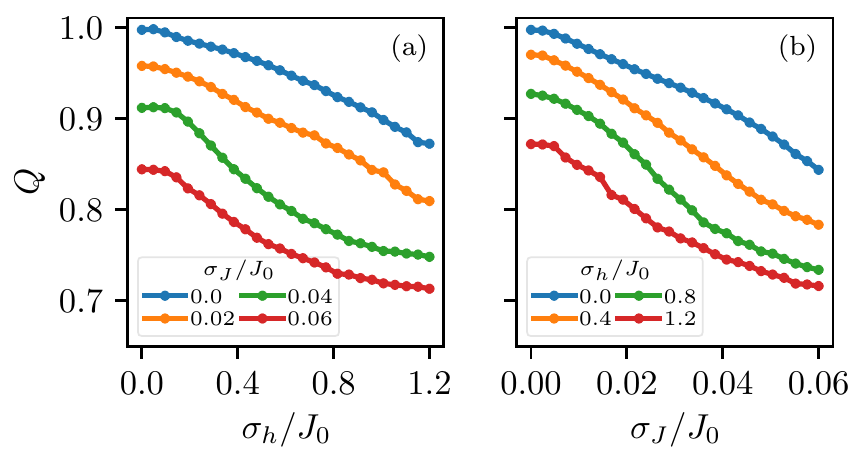}%
\caption{\label{fig:q-cut}
Quality factor $Q$
(a) as a function of the field noise $\sigma_h$ for various values of the 
charge noise $\sigma_J$,
and (b) as a function of $\sigma_J$ for various values of $\sigma_h$.
Both panels have magnetic field gradient $h_0=0$.
}
\end{figure}

\begin{figure}[]
\centering
\includegraphics{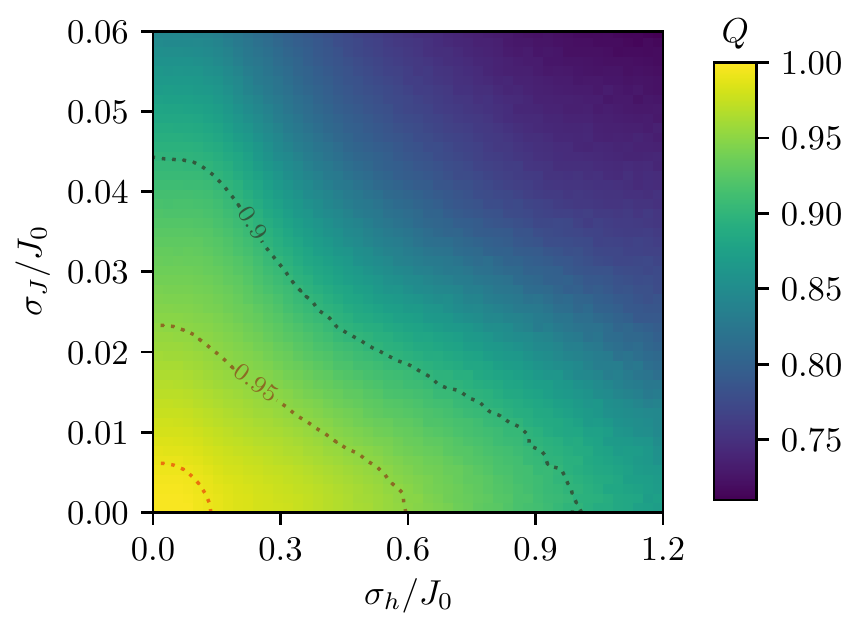}%
\caption{\label{fig:q-heatmap}
Quality factor $Q$ as a function of the charge noise $\sigma_J$ and the 
field noise $\sigma_h$,
for magnetic field gradient $h_0=0$.
The three dotted contour lines mark the levels $Q=0.9$, $0.95$, and $0.99$.
}
\end{figure}

\subsection{Noise dependence}
\label{sec:q}

We first consider the case where the magnetic field gradient is zero on 
average, $h_0 = 0$,
and examine the variation of the coherence time with respect to both the 
charge noise $\sigma_J$ and the field noise $\sigma_h$.
Within the $h_0=0$ parameter subspace, we find that the singlet-triplet qubits 
behave similarly to the exchange-only qubits as reported in 
Ref.~\onlinecite{Throckmorton17}.

Figures~\ref{fig:q-cut} and~\ref{fig:q-heatmap} show the dependence of the 
quality factor $Q$ on the charge noise $\sigma_J$ as well as 
the field noise $\sigma_h$.
We find that the quality factor for the coupled qubits is suppressed when 
either type of noise increases, and the system is significantly more 
susceptible to charge noise than field noise.
As marked by a contour line in Fig.~\ref{fig:q-heatmap},
to achieve a quality factor $Q$ higher than $0.9$ (corresponding to a 
coherence time $T_2^*\sim 10J_0^{-1}$), the maximum allowed charge noise 
$\sigma_J$ is around $0.045J_0$, while the maximum allowed field noise 
$\sigma_h$ is around $1.0J_0$.

The order of magnitude difference between the sensitivity to charge noise and 
the sensitivity to field noise as measured by two-qubit coherence time
is in agreement with the results previous reported on the exchange-only 
qubits~\cite{Throckmorton17}.
Quantitatively, we find that the singlet-triplet system is about 3 (1.5) times 
more sensitive to the charge (field) noise compared with the exchange-only 
system in the regime with a quality factor $Q\ge 0.9$.
This is consistent with the intuitive observation that the exchange-only 
system enjoys an additional protection due to the spin $S_z$ conservation.
The fact that charge noise is the dominant decohering mechanism for 
singlet-triplet qubits (and is even more detrimental here than for exchange-only
qubits) is, however, only true for $h_0=0$ as we discuss next.

\subsection{Effect of the magnetic field gradient}

\begin{figure}[]
\centering
\includegraphics{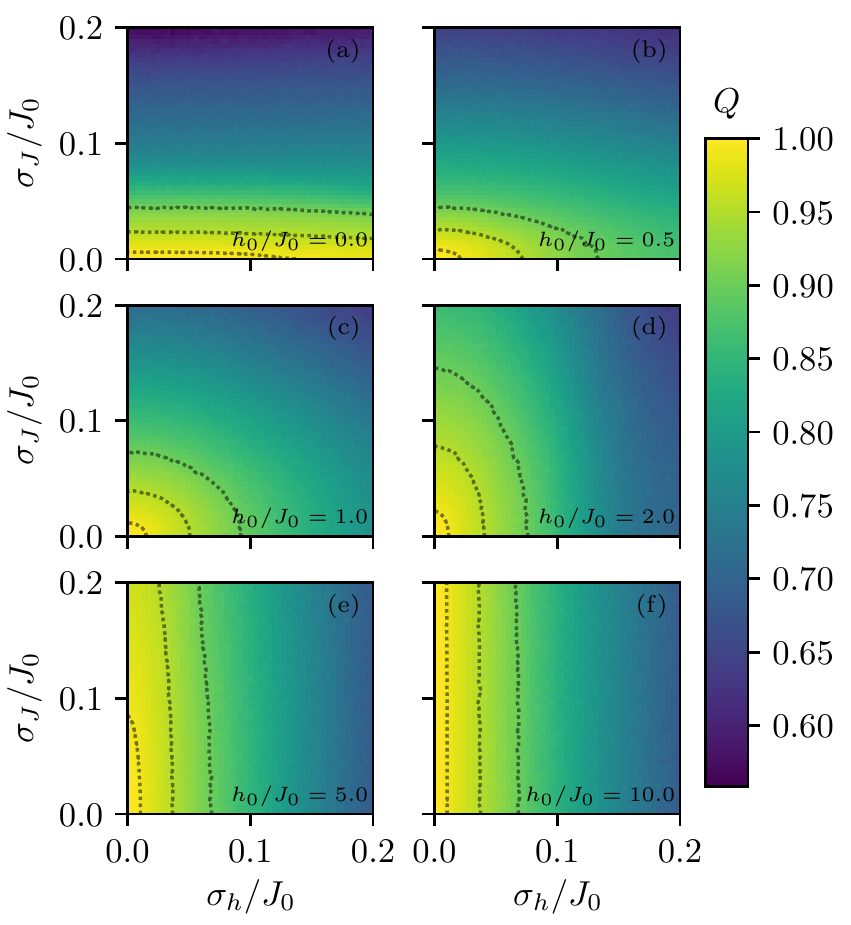}%
\caption{\label{fig:q-heatmap-h0}
Quality factor $Q$ as a function of the charge noise $\sigma_J$ and the 
field noise $\sigma_h$,
for a range of magnetic field gradient $h_0$ from $0$ to $10J_0$.
All panels share the same color map depicted on the right.
In each panel, the three dotted contour lines mark the levels 
$Q=0.99, 0.95, 0.90$ (from the bottom-left corner inwards).
}
\end{figure}

Experimentally, charge noise in GaAs-based spin-qubit devices is typically 
much weaker in absolute strength than field noise,
due to the strong Overhauser nuclear spin fluctuations.
In Si-based spin-qubit devices, however, the nuclear spin fluctuations can be 
significantly suppressed thanks to isotope purification of $^{28}$Si~\cite{Witzel10}.
In this case, the strong sensitivity to charge noise may pose a serious 
obstacle to the fidelity of coupled qubits,
since there is no known way to systematically reduce the charge noise in 
semiconductor structures.
From our numerical results, we find that this problem may be alleviated 
through the additional tunable parameter in the singlet-triplet system, namely,
the average magnetic field gradient $h_0$ across each singlet-triplet qubit.

Figure~\ref{fig:q-heatmap-h0} shows the effect of $h_0$ on the noise 
dependence of the quality factor $Q$.
As the magnetic field gradient $h_0$ goes up, the coherence of the coupled-qubit 
dynamics becomes less sensitive to the charge noise $\sigma_J$, but more 
vulnerable to the field noise $\sigma_h$.
When $h_0$ is higher than $J_0$, the coupled qubits become more susceptible to 
field noise than charge noise, in sharp contrast to the situation for 
$h_0=0$.
The sensitivity to $\sigma_h$ saturates when $h_0$ is more than a few times 
stronger than $J_0$.
For reference, we note that the entangling gate experiments reported in 
Ref.~\onlinecite{Nichol17} were carried out at an effective $h_0\sim 5J_0$, 
albeit under a different setup with individual qubits driven by an oscillatory 
$J_i(t)$.
Comparing the numerical results in Figs.~\ref{fig:q-heatmap} 
and~\ref{fig:q-heatmap-h0}(e), we find that the maximum allowed charge noise 
to achieve a high quality factor $Q\ge 0.99$
increases by more than 10 times as the magnetic field gradient $h_0$ is cranked up 
from zero to $5J_0$.
This enhanced stability against charge noise is consistent with the 
experimental observation in Ref.~\onlinecite{Nichol17} that a magnetic field
gradient $h_0\sim 5J_0$ increases the two-qubit coherence time by an order 
of magnitude in a device dominated by charge noise.
We mention here that the GaAs system used in Ref.~\onlinecite{Nichol17} obviously 
also has considerable field noise, arising from nuclear spin fluctuations in 
Ga and As, contributing to decoherence.

\section{Fidelity of entangled state preparation}
\label{sec:ent}

The two-qubit coherence time $T_2^*$ measures the persistence of two-qubit 
oscillations in the presence of environmental noises. This is a 
characterization of how well the system retains the initial non-eigenstate 
information.
In this section, we study a different aspect of two-qubit fidelity, namely, 
the fidelity $F_E$ of preparing an entangled state.
We investigate how well the system produces an entangled state starting from 
an initial product state under the influence of environmental noises.
This analysis is less sophisticated than a full-blown gate fidelity calculation
using randomized benchmarking. 
Nevertheless, it provides useful insights through a perspective complementary 
to the two-qubit coherence time $T_2^*$,
and provides a single fidelity number (the entanglement fidelity, $F_E$)
similar to the full-blown numerically intensive Clifford gate randomized 
benchmarking calculation which is beyond the scope of the current work.

\subsection{Producing an entangled state}

We choose the product initial state 
$\ket{\phi(0)}=\ket{\uparrow}_x\otimes\ket{\downarrow}_x$ and let the system 
evolve under the Hamiltonian $H$ in Eq.~\eqref{eq:h} for a fixed amount of 
time $t$ (to be specified). In the clean limit, the resulting state is
\begin{equation}
\ket{E(t)}=
e^{-i[
\varepsilon J_0^2\sigma_1^z\sigma_2^z
+J_0(\sigma_1^z+\sigma_2^z)
+h_0(\sigma_1^x+\sigma_2^x)
]t}
|\phi(0)\rangle.
\end{equation}
This is in general an entangled state, and as we discuss below, with a proper 
choice of the evolution time $t$, $\ket{E(t)}$ is in fact maximally entangled 
for both $h_0=0$ and $h_0\gg J_0$.
It should be noted that the Hamiltonian $H$ is not always effective at 
generating entanglement starting from an arbitrary initial state. The 
particular initial state 
$\ket{\phi(0)}=\ket{\uparrow}_x\otimes\ket{\downarrow}_x$ chosen here
provides a simple setup to discuss the effect of noise on entangled state 
preparation.

In the presence of environmental noises,
the time-evolved state $e^{-iHt}\ket{\phi(0)}$ depends on the disorder 
realization and deviates from its clean limit $\ket{E(t)}$.
Using the latter as a reference, we define the fidelity of entangled state 
preparation~\cite{Nelsen00} $F_E$ as the disorder-averaged overlap
\begin{equation}\label{eq:FE}
F_E=\average{\Big|\braket{E(t)|e^{-iHt}|\phi(0)}\Big|^2}.
\end{equation}
Similarly to the two-qubit coherence time $T_2^*$, this is a function of the 
field noise $\sigma_h$, the charge noise $\sigma_J$, and the magnetic field gradient 
$h_0$.

\begin{figure}[]
\centering
\includegraphics{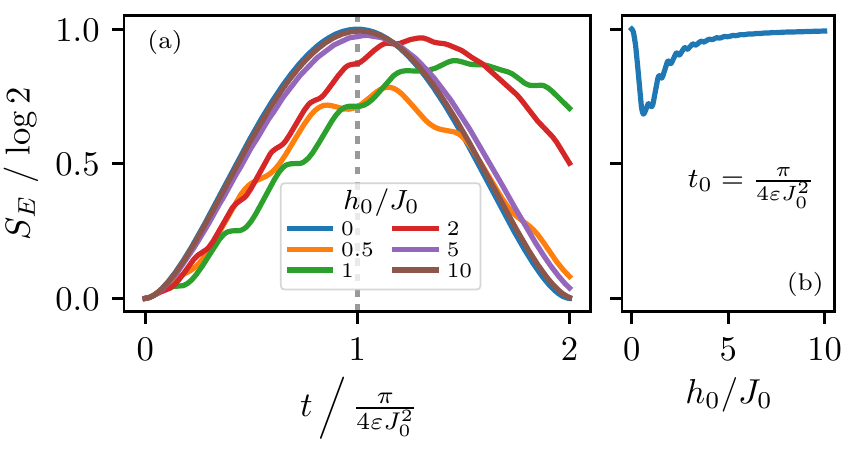}%
\caption{\label{fig:entanglement}
(a) Entanglement entropy $S_E$ of $\ket{E(t)}$ as 
a function of the evolution time $t$ for various values of $h_0$.
The evolution time $t$ is shown in units of 
$t_0=\frac{\pi}{4\varepsilon J_0^2}$, whereas the entanglement entropy 
$S_E$ is shown in units of its maximal value $\log 2$.
(b) Entanglement entropy of $\ket{E(t_0)}$ for 
$t_0=\frac{\pi}{4\varepsilon J_0^2}$, as a function of $h_0/J_0$.
Data in both panels are computed in the clean limit $\sigma_J=\sigma_h=0$.
}
\end{figure}

We want to make sure that $F_E$ indeed measures the fidelity associated with 
generating an \emph{entangled} state.
To this end, we now discuss the choice of the evolution time $t$ that 
maximizes the entanglement between the two qubits in the reference 
state $\ket{E(t)}$.
In the absence of a magnetic field gradient $h_0$, the reduced density matrix 
after tracing out one qubit in $\ket{E(t)}$ takes the simple form
\begin{equation}
\frac{1}{2}\left(\begin{array}{cc}
1 & -e^{-2i J_0t}\cos (2\varepsilon J_0^2 t) \\
-e^{2i J_0t}\cos (2\varepsilon J_0^2 t) & 1 \\
\end{array}\right).
\end{equation}
This suggests setting the evolution time $t$ in Eq.~\eqref{eq:FE} to
\begin{equation}\label{eq:t0}
t_0=\frac{\pi}{4\varepsilon J_0^2},
\end{equation}
where $\varepsilon J_0$ measures the strength of the interqubit Ising 
coupling (set to $0.1$ in this paper).
This choice ensures that the reference state $\ket{E(t_0)}$ is maximally 
entangled between the two qubits for $h_0=0$, with entanglement entropy 
$S_E=\log 2$.

The situation for $h_0\neq 0$ is less obvious.
Figure~\ref{fig:entanglement}(a) shows the dependence of the entanglement 
entropy $S_E$ between the two qubits on the evolution time $t$, for various 
values of $h_0$.
We find that for both $h_0=0$ and $h_0\gg J_0$,
the entanglement entropy of $\ket{E(t)}$ peaks at $t=t_0$, whereas for 
intermediate $h_0\sim J_0$, the entanglement entropy has irregular dynamics 
but still reaches a moderate level at $t=t_0$.
Figure~\ref{fig:entanglement}(b) shows the entanglement entropy at $t=t_0$ as 
a function of the magnetic field gradient $h_0$.
We find that the reference state at $t=t_0$ is nearly maximally entangled for 
a wide range of $h_0$ except for a small window near $J_0$.
This justifies our operational definition of the entanglement fidelity using 
the evolution time $t_0$ defined in Eq.~\eqref{eq:t0}.

\subsection{Noise dependence of $F_E$}

\begin{figure}[]
\centering
\includegraphics{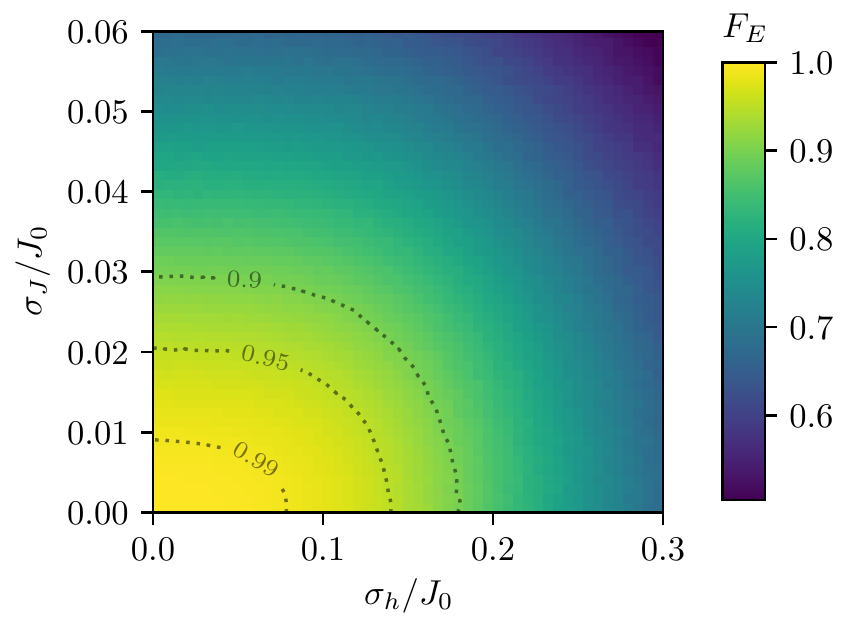}%
\caption{\label{fig:entfid-heatmap}
Entanglement fidelity $F_E$ as a function of the charge noise $\sigma_J$ and 
the field noise $\sigma_h$ at zero magnetic field gradient $h_0$.
The three dotted contour lines mark the levels $F_E=0.9$, $0.95$, and $0.99$, 
resp.
}
\end{figure}

\begin{figure}[]
\centering
\includegraphics{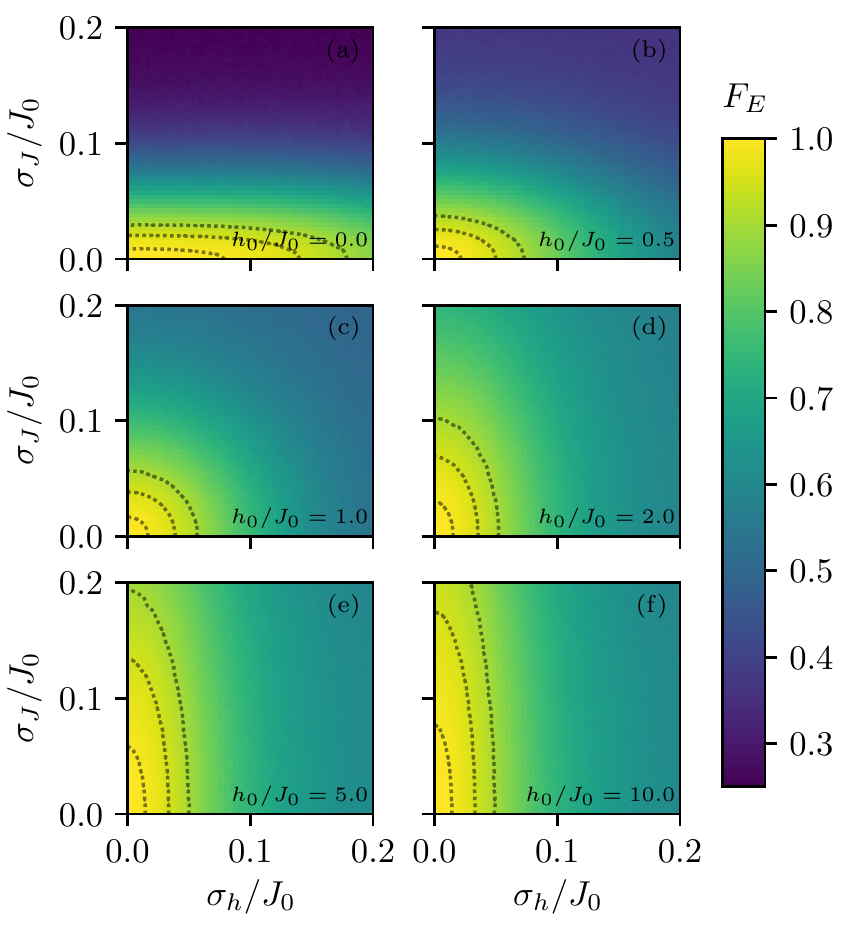}%
\caption{\label{fig:entfid-heatmap-h0}
Entanglement fidelity $F_E$ as a function of the charge 
noise $\sigma_J$ and the field noise $\sigma_h$, for a range of magnetic 
field gradient $h_0$ from $0$ to $10J_0$.
All panels share the same color map depicted on the right.
In each panel, the dotted contour lines mark the levels $F_E=0.99, 0.95, 0.90$ 
(from the bottom-left corner inwards).
}
\end{figure}

We now examine how the fidelity of entangled state preparation is affected by 
environmental noises.
First we consider the case of zero magnetic gradient $h_0=0$.
Figure~\ref{fig:entfid-heatmap} shows the dependence of the entanglement fidelity $F_E$ as 
a function of the charge noise $\sigma_J$ and the field noise $\sigma_h$.
We find that $F_E$ decays monotonically when either type of noise 
increases, and the system is more susceptible to the charge 
noise $\sigma_J$ than the field noise $\sigma_h$.
To reach $F_E$ higher than $0.9$, the maximum allowed charge noise $\sigma_J$
is around $0.03J_0$, while the maximum allowed field noise $\sigma_h$ is 
around $0.18J_0$.
We observe that the fidelity of entangled state generation has a charge noise 
dependence comparable to that of the quality factor $Q$ associated with 
the two-qubit coherence time $T_2^*$, but it has a field noise dependence 
about 5 times stronger than that of the quality factor $Q$.
This suggests that field noise is more effective at disrupting the 
precise preparing of an entangled state than damping the coupled-qubit 
oscillations.
This is germane for future progress in the subject since field noise can 
essentially be eliminated is Si qubits through isotopic purification.

Compared with the exchange-only qubits studied in 
Ref.~\onlinecite{Throckmorton17}, the entanglement fidelity $F_E$ for the singlet-triplet 
qubits computed here is significantly more susceptible to charge noise.
Intuitively, this is consistent with the fact that the Ising Hamiltonian of 
the singlet-triplet system has a weaker (by a factor of $\varepsilon J_0$) 
interqubit coupling and thus is less effective at entangling the two qubits 
than the Heisenberg Hamiltonian for the exchange-only system. The longer 
evolution time strengthens the effect of charge noise as it modifies the 
qubit precession frequency.
This particular damaging aspect of charge noise can be partially rectified by 
having stronger interqubit coupling through careful qubit geometry engineering.

The noise dependence of $F_E$ changes qualitatively when we turn on the 
magnetic field gradient $h_0$.
The progression is shown in Fig.~\ref{fig:entfid-heatmap-h0}.
As the magnetic field gradient $h_0$ increases, the entanglement fidelity $F_E$ 
quickly develops 
more sensitivity to the field noise $\sigma_h$ while becoming less susceptible to 
the charge noise $\sigma_J$.
At the turning point $h_0=J_0$, the noise dependence of $F_E$ is approximately 
symmetric with respect to $\sigma_J$ and $\sigma_h$.
As the magnetic field gradient $h_0$ increases further, the sensitivity to
charge noise is quickly suppressed, while the sensitivity to
field noise reaches a plateau.
For $h_0\gg J_0$, the entanglement fidelity $F_E$ is limited mainly by the field noise 
$\sigma_h$ (again implying a considerable advantage for isotopically purified 
Si qubits).
Overall, we find that the fidelity of entangled state preparation has a 
noise dependence qualitatively similar to that of the coherence time quality 
factor.

\section{Conclusion}
\label{sec:conclusion}

In this paper we have studied the decoherence of two singlet-triplet 
spin qubits with capacitive coupling under the influence of 
quasistatic environmental noises.
We consider two complementary decoherence measures for coupled qubits, namely, 
the two-qubit coherence time characterizing the persistence of coupled-qubit 
oscillations, and the fidelity of entangled state preparation.
Through numerical calculations, we provide a quantitative map of the 
dependence of each decoherence measure on charge noise, field noise, and 
the intraqubit magnetic field gradient.

We find that the noise dependence of the coupled-qubit coherence changes 
qualitatively as the magnetic field gradient increases.
When the (average) magnetic field gradient vanishes, the coupled-qubit system 
is more susceptible to charge noise than field noise.
For the two-qubit coherence time to be longer 
than $10J_0^{-1}$, the maximum allowed charge noise is an order of magnitude 
lower than the maximum allowed field noise. The fidelity of entangled state 
preparation has a similar (although less pronounced) bias in its noise 
sensitivity.
In contrast, when the coupled-qubit system is dominated by a strong magnetic 
field gradient, the sensitivity to charge noise is strongly suppressed and becomes 
much weaker than the sensitivity to field noise, as visible in both the two-qubit 
coherence time and the entanglement fidelity.

Our results highlight the impact of the magnetic field gradient on the noise 
dependence of the coupled-qubit system.
Increasing the magnetic field gradient $h_0$ proves to be an effective measure to 
protect against charge noise the coherence of coupled singlet-triplet qubits 
in terms of both the persistence of coupled-qubit oscillations and the precise 
preparation of entangled states.
In addition, our work points to clear advantages for Si-based qubits over GaAs 
qubits since isotopic purification could eliminate field noise in Si (but not 
in GaAs).
Elimination of field noise would enhance fidelity, and working in a large 
field gradient would suppress the charge noise, eventually leading to 
high-fidelity singlet-triplet semiconductor spin qubits suitable for quantum 
error correction protocols.
Our work establishes, however, that even in the best possible circumstances 
(Si qubits with no field noise working at a large field gradient), the magnitude 
of the effective charge noise still must be reduced below $1$-$2\%$ of the 
basic intraqubit exchange coupling $J_0$ producing the singlet-triplet qubits, 
so that a quality factor and an entanglement fidelity surpassing $99\%$ can be 
achieved for 2-qubit operations.

\acknowledgements
This work is supported by Laboratory for Physical Sciences.

\end{document}